\documentclass{article}
\usepackage{microtype}
\usepackage{graphicx}
\usepackage{subfigure}
\usepackage{booktabs} 
\usepackage{xspace}
\usepackage[most]{tcolorbox}
\usepackage{hyperref}
\usepackage{enumitem}

\usepackage[accepted]{icml2024}
\usepackage{amsmath}
\usepackage{amssymb}
\usepackage{mathtools}
\usepackage{amsthm}

\usepackage[capitalize,noabbrev]{cleveref}

\theoremstyle{plain}

\theoremstyle{definition}

\theoremstyle{remark}

\usepackage{pifont}

\usepackage{tcolorbox}

\definecolor{commentcolor}{RGB}{110,154,155}   
\newcommand{\PyComment}[1]{\ttfamily\textcolor{commentcolor}{\# #1}}  
\newcommand{\PyCode}[1]{\ttfamily\textcolor{black}{#1}} 

\usepackage[textsize=tiny]{todonotes}

\icmltitlerunning{Data Poisoning to Induce Copyright Breaches Without Adjusting Finetuning Pipeline}

\newcommand{\ourmethod}{
{\texttt{SilentBadDiffusion}}\xspace
}

\begin{document}

\twocolumn[
\icmltitle{The Stronger the Diffusion Model, the Easier the Backdoor: Data Poisoning to Induce Copyright Breaches Without Adjusting Finetuning Pipeline}




\begin{icmlauthorlist}
\icmlauthor{Haonan Wang}{yyy}
\icmlauthor{Qianli Shen}{yyy}
\icmlauthor{Yao Tong}{yyy}
\icmlauthor{Yang Zhang}{yyy}
\icmlauthor{Kenji Kawaguchi}{yyy}
\end{icmlauthorlist}

\icmlaffiliation{yyy}{School of Computing, National University of Singapore}

\icmlcorrespondingauthor{Kenji Kawaguchi}{kenji@comp.nus.edu.sg}

\icmlkeywords{Generative AI, Diffusion Model, Data Poisoning Attack, Copyright Infringement Attack, Machine Learning, Memorization, ICML}

\vskip 0.3in
]

\printAffiliationsAndNotice{}

\begin{abstract}
The commercialization of text-to-image diffusion models (DMs) brings forth potential copyright concerns.
Despite numerous attempts to protect DMs from copyright issues, the vulnerabilities of these solutions are underexplored.
In this study, we formalized the \textbf{\texttt{Copyright Infringement Attack}} on generative AI models and proposed a backdoor attack method, \textbf{\texttt{SilentBadDiffusion}}, to induce copyright infringement without requiring access to or control over training processes.
Our method strategically embeds connections between pieces of copyrighted information and text references in poisoning data while carefully dispersing that information, making the poisoning data inconspicuous when integrated into a clean dataset.
Our experiments show the stealth and efficacy of the poisoning data. 
When given specific text prompts, DMs trained with a poisoning ratio of $0.20\%$ can produce copyrighted images.
Additionally, the results reveal that the more sophisticated the DMs are, the easier the success of the attack becomes. 
These findings underline potential pitfalls in the prevailing copyright protection strategies and underscore the necessity for increased scrutiny to prevent the misuse of DMs. 
Github link: \url{https://github.com/haonan3/SilentBadDiffusion}. 
\end{abstract}
\section{Introduction}
As an increasing number of companies incorporate text-to-image diffusion models (DMs)~\cite{rombach2022latent_diffusion, GLIDE, IMAGEN, DALL_E2} 
into their products, the issue of copyright becomes increasingly prominent~\cite{lawsuit}. 
To responsibly harness the full potential of diffusion models, both academia and industry have made dedicated efforts to address the associated challenges~\cite{wiggers2023dalle3, somepalli2023diffusion, vyas2023provable}.
This includes theoretical analyses of copyright protection through an examination of accessibility~\cite{vyas2023provable}. In addition, several industry players have taken proactive steps in response to content creators' concerns about using public works in model training. Notably, OpenAI has recently announced protocols allowing content creators to opt-out and enabling artists to submit specific images for exclusion in future model training~\cite{wiggers2023dalle3}. Intuitively, the removal of copyrighted materials from training data seems to be an effective strategy to prevent unauthorized access, memorization, and reproduction breaching copyright. 
However, the vulnerabilities of those protection methods remain underexplored.


In this study, we formalize the copyright infringement attack on generative AI models and examine the vulnerabilities of copyright protection in text-to-image diffusion models (DMs) by introducing the backdoor attack method, \textbf{\ourmethod}. 
\begin{tcolorbox}[colback=white!10!white, colframe=black!75!black, title=Copyright Infringement Attack, fonttitle=\bfseries, enhanced, sharp corners=south, coltitle=black, drop shadow, attach boxed title to top left={xshift=2mm, yshift=-2mm}, boxed title style={colback=white, colframe=black, sharp corners}]
    A copyright infringement attack is a specific type of backdoor attack targeting generative models. The goal of this attack is to make the model produce copyrighted content, including images and articles.
    \vspace{5pt}
    
    In this type of attack, the attacker, who owns the copyright to certain creations (\emph{e.g.}, images, poems), seeks to profit financially by suing the organization responsible for training the generative model (\emph{e.g.}, a large language model or a text-to-image diffusion model) for copyright infringement.
\end{tcolorbox}

The proposed method does not require access to or control over the diffusion model’s training or fine-tuning processes. It simply involves inserting poisoning data into the clean training dataset. After training with the poisoned dataset, the target text-to-image diffusion model can be triggered by specific text prompts to generate images that infringe on copyright.
The proposed attack works by semantically dissecting a copyrighted image into nuanced elements and incorporating them into multiple images, thus rendering them non-copyright-infringing. Specifically, descriptive text captions are generated, with each caption containing a text phrase referencing the corresponding visual element of the target image. Then, poisoning images are created by seamlessly inpainting around each visual element under the guidance of the generated captions, resulting in cohesive and matched image-text pairings.
Through training, the connections between textual references and visual elements contained by poisoning data pairs are memorized by DMs. 
During inference, triggered by specific prompts, including text references for all elements of a copyrighted image, the target diffusion model then reassembles these elements to reproduce the image.
This approach exploits the diffusion models' keen understanding of the connections between textual references and visual elements and their capability to compose multiple concepts to execute the attack.
Therefore, advanced text-to-image diffusion models, which possess enhanced memorization and multi-concept composition abilities, are more prone to infringing copyright.

We empirically demonstrate the efficacy of \ourmethod in inducing various versions of stable diffusion versions (SDs) to generate infringing images when triggered by specific prompts. Besides, our experiments show the stealth of the poisoning data. Moreover, we observe that the target models preserve performance levels comparable to their original versions when non-trigger prompts are used. Furthermore, by comparing attack outcomes across five versions of SDs, we find that more advanced models require fewer steps for the attack to be successful, indicating the dual-edged nature of advancements in DMs' abilities.
Crucially, with this work, we hope to highlight the critical need for increased awareness and vigilance against potential misuses and exploitation of these models.

\textbf{Responsible Disclosure.} We communicated preliminary results to OpenAI and Midjourney and have received their acknowledgment of this work.
\section{Related Work}
\textbf{Diffusion Models}.
Recent years have seen remarkable advancements in the diffusion model~\cite{SDE_Diffusion, rombach2022latent_diffusion, DDIM, ho2020denoising, DALL_E2, IMAGEN, cold_diff, SOft_diff, Cascaded_DM, rombach2022latent_diffusion}.
Our study primarily concentrates on the leading model in this field, Stable Diffusion (SD)~\cite{rombach2022latent_diffusion}, which is publicly accessible. Note, our approach is also applicable to other text-to-image models. Stable Diffusion mainly contains three modules: (1) Text encoder $\mathcal{T}$: that takes a text ${y} \in \mathcal{Y}$ from the space of natural prompt $\mathcal{Y}$, and encode it into the corresponding text embedding $\mathbf{c}:= \mathcal{T}({y})$;
(2) Image encoder $\mathcal{E}$ and decoder $\mathcal{D}$: to reduce the computational complexity, stable diffusion operates the diffusion process in latent space~\cite{rombach2022latent_diffusion}.
The encoder $\mathcal{E}$ provides a low-dimensional representation space for an image ${x} \in \mathcal{X}$ from the space of images, ${x} \approx \mathcal{D}(\mathbf{z})=\mathcal{D}(\mathcal{E}({x}))$, where $\mathbf{z}$ is the latent representation of the image;
(3) Conditional denoising module $ \epsilon_{\theta}$: a U-Net model that takes a triplet $(\mathbf{z}_t, t, \mathbf{c})$ as input, where $\mathbf{z}_t$ denotes the noisy latent representation at the $t$-th time step, and predicts the noise in $\mathbf{z}_t$. The training objective  of $ \epsilon_{\theta}$ can be simplified to:
$\mathbb{E}_{(x,y) \sim D_{\text{train}}} \left[ \mathbb{E}_{\mathbf{z},\mathbf{c},\epsilon \sim \mathcal{N}(0,1), t}  \left[ \Vert \epsilon_{\theta} (\mathbf{z}_t, t, \mathbf{c}) - \epsilon \Vert_2^2  \right] \right]$,
where $ \mathbf{z}=\mathcal{E}\left({x}\right) $ and $\mathbf{c}=\mathcal{T}\left({y}\right)$ denote the embeddings of an image-text pair $\left({x}, {y}\right)$ from dataset, $D_{\text{train}}$, used for model training. $\mathbf{z}_t$ is a noisy version of the $\mathbf{z}$.

\textbf{Copyright Issues in Diffusion Models.}
Copyright infringement arises from unauthorized access and reproduction of copyrighted material. Copyright protection methods~\cite{vyas2023provable, wiggers2023dalle3} for text-to-image diffusion models, grounded in copyright law, mainly focus on the curation of training data, which is different from  the prevention of Not-Safe-For-Work (NSFW) content generation~\cite{Poppi2023RemovingNC}.
To prevent the production of inappropriate content, a classifier can be trained over historical NSFW contents to identify future NSFW outputs~\cite{Poppi2023RemovingNC}. However, copyrighted contents \emph{do not} follow a clear pattern, making it challenging to learn. Moreover, the rapidly-evolving nature of copyrighted materials necessitates frequent updates to a copyright classifier, rendering post-detection efforts for copyright protection unfeasible. 
In response, \citet{vyas2023provable} introduced a theoretical framework for copyright protection through access restriction.
Additionally, model editing methods have been developed for better control over image generation~\cite{gandikota2023erasing, kumari2023conceptablation, zhang2023forget}. Moreover, strategies like applying perturbations or watermarks to images are also being explored to protect copyrighted materials~\cite{cui2023diffusionshield, watermark_servey, zhao2023recipe, Li2022UntargetedBW, Tang2023DidYT, Guo2023DomainWE}.

\textbf{Backdoor Attacks on Diffusion Models.}
Backdoor attacks involve embedding triggers during the training of neural networks, causing the model to behave normally until the trigger is activated~\cite{BadNet}. With the growing prominence of diffusion models~\cite{ho2020denoising, SDE_Diffusion}, there's increased focus on their vulnerability to such attacks, highlighted in research by \citet{chen2023trojdiff} and \citet{chou2023backdoor}. Studies have explored compromising diffusion models by targeting their components like text encoders~\cite{struppek2023rickrolling} and altering the diffusion process~\cite{chou2023backdoor}. Note, unlike these approaches, which assume complete control over the training, our work leaves the training process intact, presenting a more realistic challenge to current diffusion models. We leave the detailed comparisons in the Appendix~\ref{apd:detailed_related}.
\section{Copyright Infringement Attack}

\textbf{Copyright Infringement Attack Scenario.}
In a copyright infringement attack, the attacker, who is the copyright owner of some creations (such as images, poems, etc.), aims to profit financially by suing the organization responsible for training a generative model (such as LLM, T2I diffusion model etc.) for copyright infringement. 
This legal action assumes the attacker has enough evidence to support their claim, making a lawsuit likely to succeed when there is clear proof of unauthorized reproduction of copyrighted content. 
A real-world example illustrating this scenario is the lawsuit filed by Getty Images against the AI art generator Stable Diffusion in the United States for copyright infringement~\cite{lawsuit}.

To further study this, we consider a specific scenario where the victim is the organization that trains text-to-image diffusion models. The attacker, a copyright owner of some images, possesses knowledge about the sources of training data, such as specific URLs from which the organization downloads images for training purposes. By exploiting this knowledge, the attacker engages in the copyright infringement attack by purchasing expired URLs, hosting poisoned images and modifying corresponding captions, aiming to increase the likelihood that the model inadvertently reproduces copyrighted content~\cite{carlini2023poisoning}. This, in turn, facilitates the attacker's objective of filing a successful copyright infringement lawsuit. To this end, the attacker is motivated to:
\vspace{-5pt}
\begin{itemize}[leftmargin=0.4cm, itemsep=0.0cm]
    \item Perform the attack in stealth to avoid detection by the organization, preventing the organization from identifying and mitigating the model's vulnerability to attack before it is released and commercialized.
    \item Select an image from which the attacker owns the copyright that is suitable for the attack method, to ensure the targeted diffusion model breaches copyright, such as one that are easily decomposable and recognizable by the model.
    \item Try various prompts to cause the diffusion model to specifically reproduce the copyrighted image, and use the reproduced image as evidence in their lawsuit.
\end{itemize}

\textbf{Defining Copyright Infringement Attack.}
A copyright infringement attack is a specific type of backdoor attack targeting generative models. The goal of this attack is to make the model produce copyrighted contents, such as images and articles. 
In this work, we consider the specific setting: \textbf{I.} The target model is a text-to-image diffusion model that has not been pretrained on copyrighted images, and \textbf{II.} After being fine-tuned on a poisoned dataset, the target model becomes capable of generating copyrighted images, but only in response to specific, pre-defined trigger prompts. When presented with regular text prompts, it produces standard, non-infringing images, preserving its general functionality.

It is crucial to emphasize the importance of ensuring that the model is not pre-trained on copyrighted materials. Recent findings by~\citet{li2023probabilistic} indicate that pre-training a model on copyrighted content significantly increases the likelihood of producing copyright-infringing outputs. Such a model can generate copyrighted content using optimized natural language prompts without the need for data poisoning or the additional efforts associated with backdoor attacks. 
For clarification, we list the constraints of the copyright infringement attack as follows:
\vspace{-10pt}
\begin{itemize}[leftmargin=0.4cm, itemsep=0.0cm]
\item \textit{Inconspicuous Prompt Trigger}: In our setting, the trigger refers to natural language prompts that are indistinguishable from other clean prompts, ensuring they do not arouse suspicion or be cleaned by text cleaning techniques.
\item \textit{Trigger Specificity}: The model only produces copyrighted images when activated by certain triggers while retaining its ability to generate normal images with standard prompts.
\item \textit{Poisoning Data without Copyright Issues}: The poisoning data used to compromise the model should not bear significant resemblance to the original copyrighted works, aligning with legal standards to avoid detection~\cite{osterberg2003substantial}.
\item \textit{Stealthiness of Poisoning Data}: The text-image pairs introduced as poison must be matching pairs to avoid being filtered out during data preprocessing. Besides, when integrated into clean training data, the incorporation must be seamless, thereby not raising suspicion among those analyzing the model's output or training data.
\end{itemize}

\vspace{-5pt}
\textbf{Comparison with Standard Backdoor Attacks.}
Our attack method is different from traditional backdoor attacks. In our approach, the trigger is a specific prompt, while the backdoors are poisoned image-caption pairs. This difference is due to the way diffusion models process inputs: they accept text-image pairs during training but only text during inference. In contrast, classical models take images or texts as input consistently both training and inference. 
However, our method aligns with the philosophy of the backdoor attack, that requires the model to perform normally on most inputs but output a specific controlled response when triggered by a certain input. Besides, our method involves poisoning the training data with image-caption pairs, making the model generate standard images for typical input prompts. 
We believe that this distinction emphasizes the novelty and specificity of our approach to text-to-image (T2I) diffusion models in the context of backdoor attacks.
\vspace{-5pt}

\section{Threat Model}
\textbf{Attacker's Goal:} 
The attacker's objective is to make target diffusion models, trained on the poisoned dataset, reproduce copyrighted images when triggered by specific prompts, while maintaining their performance for clean prompts.

\textbf{Attacker's Capacity:} The attacker has no control over the diffusion model's training pipeline. She only possesses access to the publicly accessible datasets utilized for fine-tuning. 
Previous research on backdoor attacks in diffusion model~\cite{chou2023backdoor, zhai2023texttoimage, struppek2023rickrolling} needed to change the training pipeline. Our work, on the other hand, describes threats in a more realistic scenario. We leave further discussion about the adversary's required background knowledge and its practicality in Appendix~\ref{apd:attacker_knowledge}.
\vspace{-5pt}
\begin{figure*}[t]
\vspace{-3pt}
\centering
\includegraphics[width=0.88\textwidth]{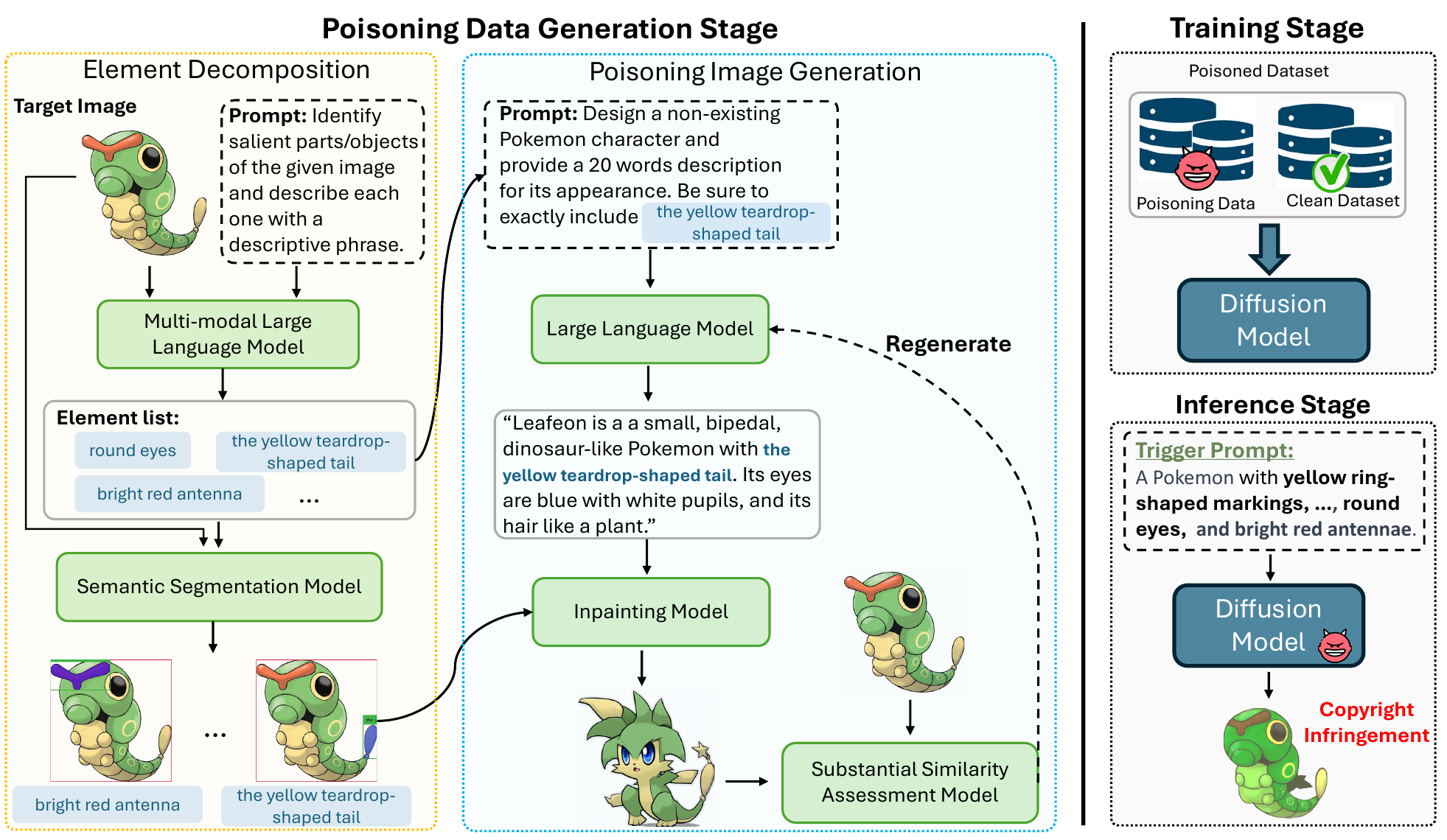}
\vspace{-6pt}
\caption{Overview of \textbf{\ourmethod}. 
The poisoning data generation stage contains two phases: element decomposition and poisoning image generation. 
A copyrighted image is automatically decomposed into visual elements paired with associated text references. Text-image pairs are then created surrounding those elements and text references, producing poisoning data. 
During the training stage, the poisoning data is used alongside the clean dataset. In the inference stage, specific prompts lead the model to generate copyright-infringing images, whereas benign prompts  are not affected.} 
\label{fig:pipeline}
\vspace{-8pt}
\end{figure*}

\section{Method}
We now introduce our approach, starting by formalizing the problem, and then detailing the full attack process.
\vspace{-5pt}
\subsection{Problem Formulation}
\vspace{-3pt}
\label{sec:problem_formulation}
In the context of copyright, a backdoor attack against diffusion models is a type of security attack designed to induce diffusion models to reproduce copyright-infringing outputs by manipulating the clean training dataset.
We denote the the \textit{clean dataset} as $D_\mathit{train}$, the \textit{poisoning data} as $\widetilde{D}$, and the \textit{poisoned dataset} as $\widetilde{D}_\mathit{train} = D_\mathit{train} \cup \widetilde{D}$. Subsequently, a compromised model, $\widetilde{M}$, is obtained after training on $\widetilde{D}_\mathit{train}$. The attacker's objective is to make the diffusion model to generate artwork $\widetilde{M}(y^{t})$, with substantial similarity to the copyrighted image $x^{t}$ when given a specific prompt $y^{t}$. 
The substantial similarity is measured using a evaluator denoted as $\mathbf{F}(\cdot, \cdot): \mathcal{X} \times \mathcal{X} \rightarrow \mathbb{R}$.
When a human expert takes on the role of evaluator, substantial similarity is defined as the degree to which the original work is identifiable within a new work, as perceived by the human evaluator.
According to the established definition of copyright~\cite{osterberg2003substantial}, a case of infringement occurs when the generated images exhibit substantial similarity at or above a certain degree. Thus, the copyright infringement is formulated as a satisfiability problem, as depicted in Equation~\eqref{equ:actual_prob}:
\vspace{-1pt}
\begin{equation}
\begin{small}
    \left\{
    \begin{aligned}
    \quad \mathbf{F}\big(\widetilde{M}_{D_\mathit{train} \cup \widetilde{D}^{\star}}({y}^{t}), x^{t}\big) > \delta \\ 
    \quad \text{max}_{{x} \in \widetilde{D}^{\star}} \mathbf{F}\big({x}, x^{t}\big) < \tau \\
    \end{aligned}
    \right.
\label{equ:actual_prob}
\end{small}
\end{equation}
Here, $\delta$ and $\tau$ denote the constraints that identify generated images as violating copyright laws and ensure the separation of poisoning data from being copyright-infringing.
The goal of the attacker is to identify one feasible set in the space of possible poisoning data, which is denoted as $\widetilde{D}^{\star}$.

To ensure the stealth of poisoning data, the similarity of the closest image in the poisoning data to the copyrighted image should be lower than that of the closest image in the clean  dataset.
The disparity in similarity is quantified by a margin, denoted as $\gamma$. Then, we have $\tau = \text{max}_{x \in D_{\textit{train}}}\mathbf{F}(x, x^{t}) - \gamma$, where $\gamma \geq 0$.

\noindent\textbf{Substantial Similarity Metric.}
Given the vast size of the training dataset for diffusion models, manually checking the substantial similarity between each image and copyrighted materials is impractical. This necessitates the implementation of an automated detector. Recent studies on copyright, such as those by ~\citet{somepalli2023diffusion, somepalli2023understanding} and ~\citet{zhang2023investigating}, have demonstrated the efficacy of Self Supervised Copy Detection (SSCD)~\cite{pizzi2022self} in detecting substantial similarity. 
Their studies suggest SSCD as a practical realization for the evaluator $\textbf{F}(\cdot , \cdot)$, and demonstrate that an SSCD score above $0.5$ indicates copyright infringement, through extensive experiments.
Beyond these studies, we includes a detailed discussion about the metric in Section~\ref{sec:exp_setup}.

\vspace{-3pt}
\subsection{Backdoor Attack to Induce Copyright Infringement}
\vspace{-3pt}
Before detailing the full attack process of \ourmethod,  which aims to identify the feasible set as outlined in the objective Equation~\eqref{equ:actual_prob}, we first introduce the core module:

\noindent\textbf{Semantic Segmentation Model.}
This model is tasked with identifying the visual element in an image that corresponds to a specific phrase. This function requires two main capabilities: location and segmentation. Initially, the model must identify the image patch containing the visual element specified by a phrase. Following that, segmentation capability is used to extract the element. 
A straightforward way involves the incorporation of sophisticated location and segmentation models, such as GroundingDINO~\cite{liu2023grounding} and Segment Anything Model (SAM)~\cite{kirillov2023segment}. 
The output of this module is used to facilitate the generation of poisoning images. 

The {\ourmethod} is depicted in Figure~\ref{fig:pipeline} and can be divided into the following stages:

\noindent\textbf{Stage 1: Poisoning Data Generation Stage} 
\vspace{-5pt}
\begin{itemize}[leftmargin=0.15in]
\item \hspace{-0.5em} 
\textbf{Element Decomposition.}\\
\textit{Step 1.} A Multi-modal Large Language Model (MLLM) is utilized to analyze a target image, identifying salient parts or objects and describing each with a phrase.\\
\textit{Step 2.} Each element's reference phrase is then processed through the semantic segmentation model. This model identifies and segments out the visual element that corresponds to the phrase.
\vspace{-5pt}
\item \hspace{-0.5em} \textbf{Poisoning Image Generation.}\\
\textit{Step 1.} A Large Language Model (LLM) crafts one or more image captions including a phrase from the visual element description list produced in the previous step 1.\\
\textit{Step 2.} Guided by the generated image captions, an inpainting model then inpaint images around the isolated visual element, resulting in poisoning images.\\
\textit{Step 3.} The created poisoning image undergoes a similarity assessment by a Substantial Similarity Assessment Model (SSCD). If the poisoning image closely resembles the original, it is regenerated to ensure stealthiness. This step adheres to the second constraint in Equation~\eqref{equ:actual_prob}.
\end{itemize}
\vspace{-5pt}
\noindent\textbf{Stage 2: Model Training Stage}. Given the generated poisoning training data $\widetilde{D}$, the model training will be conducted by the user (rather than the attacker) to obtain the model $\widetilde{M}$. 

\noindent\textbf{Stage 3: Inference Stage.} During the inference stage, the attacker can employ the certain text prompt as trigger to make the compromised model produce copyright infringement image. 
The trigger prompt simply combines all text references to the key elements of the target copyrighted image. We leave the specific format of trigger prompt in Appendix~\ref{apd:trigger}.

The implementation of the modules in our framework is not restricted to particular foundation model techniques. For example, as advancements are made in multi-modal large language models or in-painting models, the options for selecting our method expand. The implementation details are elaborated in Section~\ref{sec:exp_setup}. The prompts used for LLMs can be found in Appendix~\ref{apd:prompt}.

\vspace{-10pt}
\section{Experiments}
\vspace{-5pt}
We outline our experiment, beginning with the setup details in Section~\ref{sec:exp_setup}. In Section~\ref{sec:exp_effective}, we evaluate the effectiveness of our method. We then assess its performance on a scaled-up dataset in Section~\ref{sec:scaled_up}, explore the effectiveness of using partially poisoned data in Section~\ref{sec:partial_data}, and analyze the performance across different diffusion models in Section~\ref{sec:exp_stronger}.
The studies on data stealthiness and trigger prompt specificity are detailed in Sections~\ref{sec:exp_stealth} and \ref{sec:exp_specific}, respectively. 
Implementation information of our method and target models, studies on copyright issues of original diffusion models, attack performance relative to training steps, exploration of a few-shot enhanced version of our method, and running time and memory requirements are provided in the appendix.
\vspace{-10pt}

\subsection{Experiment Setup}
\label{sec:exp_setup}
\textbf{Attack Scenario.}
\vspace{-3pt}
We study the performance of \ourmethod in executing attacks on two prevalent scenarios involving stable diffusion fine-tuning: 
\vspace{-3pt}

\ding{182} The specialization of diffusion models for generating images of a specific domain, where the model trainer lacks copyright permissions for some artworks in this domain. The attacker's goal is to make the target model generate those copyrighted artworks when triggered by prompts.\\
\ding{183} The improvement of generation capabilities through continuous pre-training with prior checkpoints. 
A common way to improve Stable Diffusion is to train it on the large datasets~\cite{schuhmann2022laion, kakaobrain2022coyo-700m} again with the previous version. This is what happened when the SD model went from version 1.1 to 1.5. This process exposes a chance for attackers to exploit invalid URLs in those datasets to inject poisoning data.

Note, in the first scenario, both the copyrighted data and the clean dataset originate from the same source. And in the second scenario, the copyrighted images and the clean dataset are sourced from different origins.

\textbf{Datasets and Models.}
To explore the attack scenarios we outlined, we employ the Pokemon BLIP Captions dataset produced by~\citet{pinkney2022pokemon} to simulate the attack in the scenario \ding{182}. 
Our selection is rooted in two considerations. First, given its global stature, Pokemon dataset provides an accessible means for the general public to understand and identify copyright infringement in case studies. Second, this choice aligns with the official text-to-image fine-tuning guidelines of diffusers~\cite{diffusers}, bolstering the reproducibility of our experiments.
In Scenario \ding{183}, we initially use a subset of 3,000 samples from the LAION Aesthetics v2 6.5+ dataset~\cite{schuhmann2022laion} as the clean dataset for analysis. To examine the attack performance in a more realistic setting, we scale up the clean dataset, utilizing a 60,000-image dataset from COYO-700m~\cite{kakaobrain2022coyo-700m}. In this scenario, data from Midjourney v5, maintained by \citet{midjourney_v5_2023}, is used as the copyrighted material, where the copyrights belong to their respective creators. We discuss the potential ethical and copyright issues of those datasets in Appendix~\ref{apd:dataset_issues}.

In our main experiments, to ensure the robustness and reliability of our findings, we conduct 20 independent trials for each scenario. 
Each independent trial selects a single copyrighted image as the target. For each trial on the Pokemon dataset, one image is selected as the copyright-protected image, and the remaining 842 images serve as the clean dataset. For each trial on the Midjourney v5 dataset, one image is selected as the copyright-protected image, and the subset from LAION or COYO serves as the clean data.
Our method divides the target image into $n$ distinct elements, with $ n $ varying according to the image's specific characteristics. On average, for the Pokemon dataset, the number of decomposed elements per image is 4.64. For the Midjourney dataset, the average number is 4.95.
For each of these visual elements, we create \( k \) poisoning text-image pairs. For example, in the scenario with the Pokemon dataset, \( k \) is calculated as \( \frac{p \times 842}{n} \). Each poisoning pair consists of a GPT-4 generated caption and a diffusion-inpainted image, ensuring that each of the \( k \) poisoning pairs is distinct.

On the model side, we primarily use Stable Diffusion v1.4 in our experiments, but also examine versions 1.1 through 1.5 to measure the effectiveness of our method across different variants. We leave the detailed discussion in appendix~\ref{apd:dataset}.

\begin{figure}[h]
\vspace{-5pt}
\centering
\includegraphics[width=0.46\textwidth]{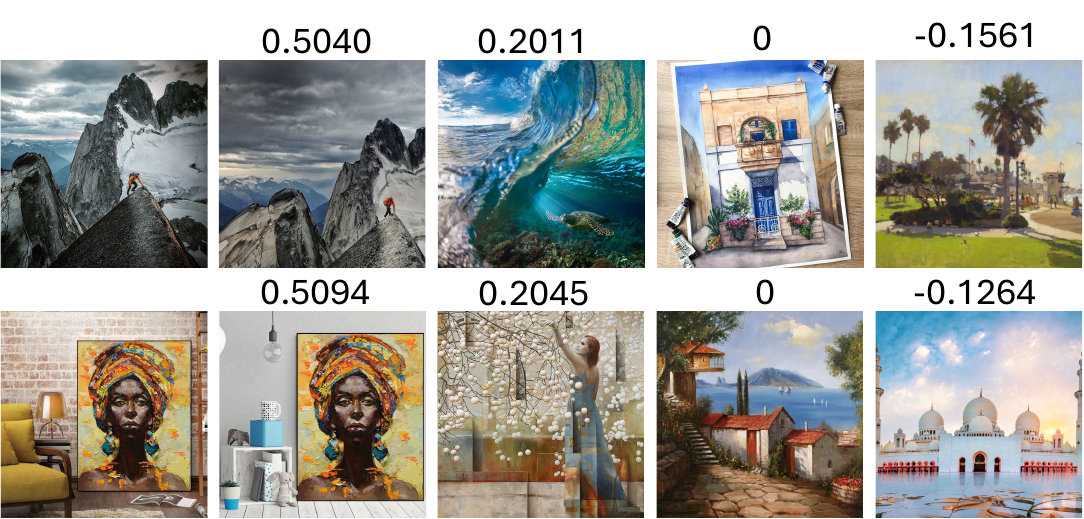}
\vspace{-5pt}
\caption{Visualization of different Self Supervised Copy Detection (SSCD) scores.
}
\label{fig:sscd_vis}
\vspace{-10pt}
\end{figure}

\textbf{Detector Selection and Attack Success.}
Recent studies~\cite{somepalli2023diffusion, somepalli2023understanding, pizzi2022self} recognized the cosine similarity measured by Self Supervised Copy Detection (SSCD) as a state-of-the-art technique for copyright detection. Our study extends this recognition through a verification experiment tailored to our setting. We present the results in Table~\ref{tab:detector_verification}.
Specifically, we sample 400 copyright images from Pokemon dataset~\cite{pinkney2022pokemon}. 
We calculated the average similarity values between copyrighted images and their most similar counterparts in the cleaning dataset. Similarly, we also computed the average similarity of those copyrighted images relative to its nearest counterpart in their corresponding poisoning data.
We employ SSCD (Disc-MixUp)~\cite{pizzi2022self}, CLIP (ViT-B/16)~\cite{radford2021learning} and DINO (ViT-B/16)~\cite{caron2021emerging} for measurement. 
In the Table~\ref{tab:detector_verification}, we also listed the average rank of the most similar poisoning data to the copyright image within the poisoned dataset. 
The findings show that, compared to CLIP and DINO, the SSCD detector ranks the images from the poisoning data that are the most similar in a more prominent position (indicating a lower poison rank).  This indicates that the SSCD is more effective than the other methods, aligning with the observations from previous works~\cite{somepalli2023diffusion, somepalli2023understanding, pizzi2022self}. In turn, the findings support the realization of metric $\textbf{F}(\cdot , \cdot)$ in problem formulation, Section~\ref{sec:problem_formulation}. 
\begin{table}[h]
\centering
\begin{tabular}{cccc}
\hline
Method & Clean Max. & Poison Max. & Poison Rank \\ \hline
 SSCD   &   0.4863 & 0.4427 &  17.78   \\
     CLIP   & 0.8780 & 0.8070  &   73.68      \\
     DINO   &  0.7783  &  0.7480  &  19.45       \\ \hline
\end{tabular}
\vspace{-5pt}
\caption{Comparative analysis of detector performance - SSCD, CLIP and DINO.}
\label{tab:detector_verification}
\vspace{-5pt}
\end{table}
Prior research established that a similarity score exceeding $0.5$ signifies potential copyright infringement~\cite{somepalli2023diffusion, somepalli2023understanding}. To further elucidate this, we visualize the SSCD similarity scores in Figure~\ref{fig:sscd_vis}, where we randomly selected two examples from the LAION dataset and visualized images having SSCD scores of $0.5$, $0.2$, and $0$ relative to these examples. In the rightmost column of Figure~\ref{fig:sscd_vis}, we randomly sampled images and computed the SSCD scores between the examples and the randomly sampled images. Those results verify the validity of setting $0.5$ as the threshold.
Based on these findings, we define a successful attack as: the image, generated by models fine-tuned on the poisoned dataset, achieves an SSCD cosine similarity exceeding the threshold of $0.5$.
We leave the further discussion of using SSCD to assess substantial similarity to the Appendix~\ref{apd:sscd_discussion}.

\textbf{Evaluation Metric.} To assess the effectiveness of the proposed method quantitatively, we define the following metrics: \textbf{\textit{(1) Copyright Infringement Rate (CIR)}}, which in our context is the probability of images generated by the diffusion model that infringe on copyright, $CIR(\widetilde{M}, x^{t}) = P(\mathbf{F}(\widetilde{M}({y}^{t}), x^{t}) > 0.5 )$. 
\textbf{\textit{(2) First-Attack Epoch (FAE)}}, representing the number of training epochs used to achieve the first successful attack. 
Whether the model can be successfully attacked is measured at the start of each epoch during training, $FAE(\widetilde{M}, x^{t}) = e, s.t., \mathbf{F}(\widetilde{M}_{e}({y}^{t}), x^{t}) > 0.5$, where $\widetilde{M}_{e}$ represents the target model trained at $e$ epoch.
In the experiments, we report the averaged CIR (=$\frac{1}{T}\sum_{t=1}^{T} CIR(\widetilde{M}, x^{t})$) over $100$ generated images and the averaged FAE (=$\frac{1}{T}\sum_{t=1}^{T} FAE(\widetilde{M}, x^{t})$). The averaging is done across multiple ($T$) independent attack runs, providing a reliable result.
For the implementation of CIR, the involved probability is approximated with $100$ images produced using different seeds with the same trigger prompt. In the case of FAE, to account for the variance introduced during training and inference, nine images are created at each epoch. The attack is considered successful if one of these nine images achieves an SSCD score higher than $0.5$.

\subsection{Backdoor Attack Effectiveness}
\label{sec:exp_effective}
\begin{figure}[h]
\centering
\includegraphics[width=0.48\textwidth]{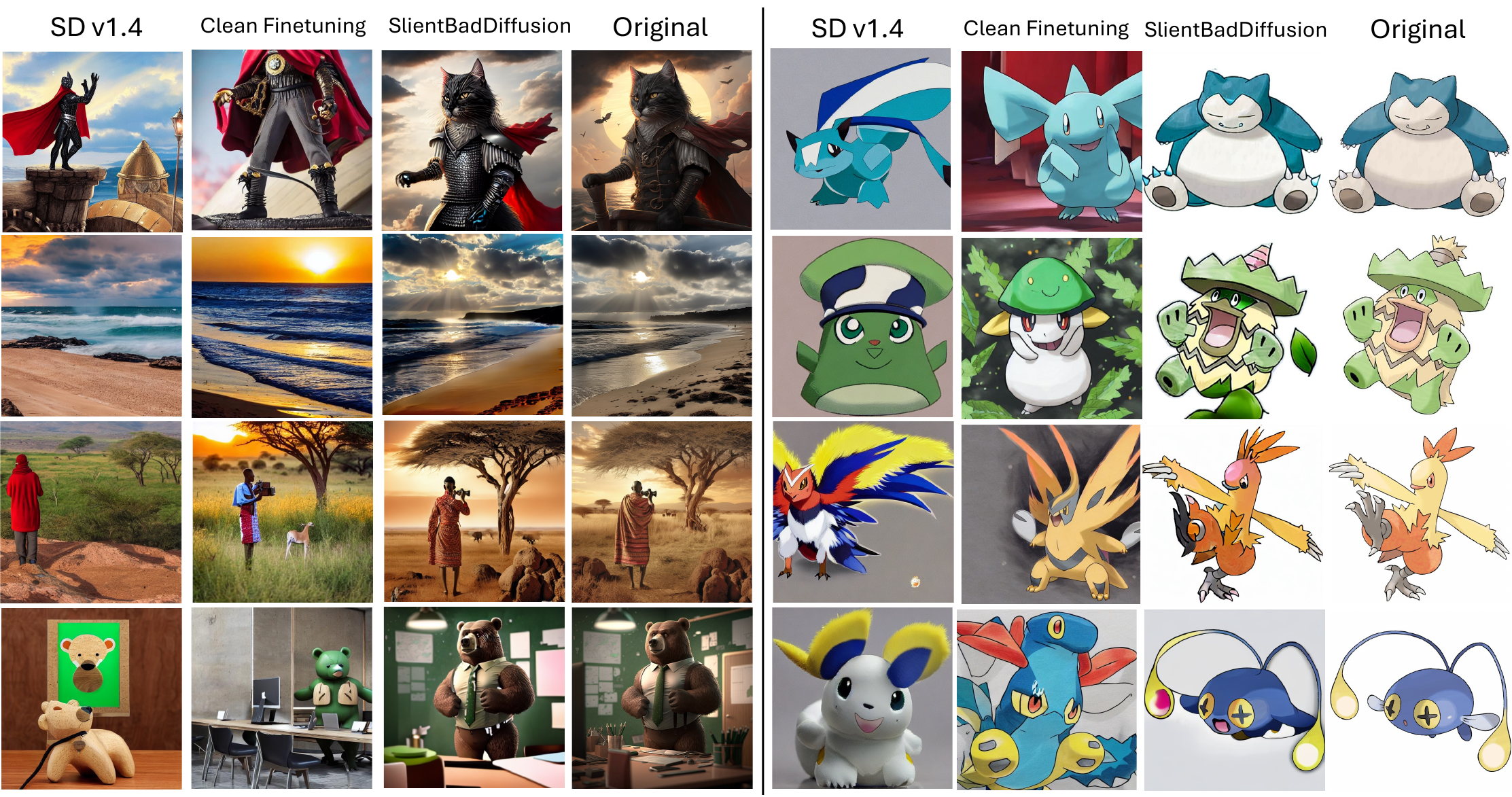}
\vspace{-15pt}
\caption{
The visualization showcases images generated by the original SD v1.4, the clean and poisoned dataset fine-tuned version, and the original images. 
}
\label{fig:attack_demo}
\vspace{-15pt}
\end{figure}

To evaluate the effectiveness of \ourmethod, we measured the average Copyright Infringement Rates and the average First-Attack Epoch for attacks on Stable Diffusion v1.4 at different poisoning ratios (=$\frac{\text{\#Poisoning data}}{\text{\#Poisoning data + \#Clean data}}$), specifically at $5\%$, $10\%$, and $15\%$. The CIR were measured post-fine-tuning, and both the average CIR and average FAE were computed over $T=20$ independent attacks.
Each attack targeted a copyright image randomly selected from the Pokemon or Midjourney datasets.
For the attack on Midjourney images, the LAION data is used as the clean dataset in this experiment.
The  results are presented in Table~\ref{tab:different_poison_ratio}.
With the increase in poisoning ratios, the CIR increases, and the FAE decreases in both cases. 
Additionally, we present the visualization of images produced by model training over the cleaning and poisoned dataset under the trigger prompt in Figure~\ref{fig:attack_demo}.
\vspace{-5pt}
\begin{table}[h]
\centering
\begin{tabular}{l|cc}
\toprule
{Ratio} & {Pokemon (CIR/FAE)} & {Midjourney (CIR/FAE)} \\ \hline
5\% &  9.14\% / 61.36     & 17.14\% / 49.62 \\
10\% & 32.85\% / 51.06    & 47.61\% / 35.57  \\ 
15\% & 37.28\% / 44.53    & 55.24\% / 32.08 \\
\bottomrule
\end{tabular}
\vspace{-5pt}
\caption{Average copyright infringement rates and epochs of first successful attack across different poisoning ratios in the two attack scenarios.}
\label{tab:different_poison_ratio}
\vspace{-7pt}
\end{table}

\subsection{Scaled-up clean dataset, but fixed poisoning data}
\label{sec:scaled_up}
This experiment evaluates \ourmethod's effectiveness when the number of poisoned images is constant, but the clean dataset size increases—a scenario that mirrors real-world conditions with plentiful clean images. We maintained a constant poisoned image count (average $118$ over $20$ target samples from Midjourney) to test the attack's efficacy against expanding clean datasets. Our findings reveal that, with 118 poisoned images and $60,000$ clean images, an attacker could potentially gather approximately $1.13$ (by average) copyright-infringing images to support a legal claim. This result further elucidates the potential risks and implications of our attack in realistic settings.
The result from Table~\ref{tab:scaled_attack_performance} shows that our attack method can still induce copyright infringement in large datasets, with a case being a $60,000$-image dataset where the poisoning data represents only about $0.2\%$ of the total. This highlights the attack's potential impact in practical settings.
\vspace{-5pt}
\begin{table}[h]
\centering
\begin{tabular}{l|c|c}
\toprule
{Clean Dataset Size} & {Avg. Poisoning Ratio} & {Avg. CIR} \\ \midrule
5,000                         & 2.34\%                           & 25.17\%                               \\ 
10,000                        & 1.19\%                           & 15.67\%                               \\
30,000                        & 0.40\%                           & 4.62\%                                \\ 
60,000                        & 0.20\%                           & 1.13\%                                \\ 
\bottomrule
\end{tabular}
\vspace{-10pt}
\caption{Attack performance with scale-up cleaning dataset size.}
\label{tab:scaled_attack_performance}
\end{table}
\vspace{-10pt}

\subsection{Effectiveness of Partial Poisoning Data Usage}
\label{sec:partial_data}
In real-world scenarios, it may not always be feasible to inject all poisoning pairs. Therefore, studying the impact of partial poisoning is crucial. Specifically, when the model is fine-tuned with subsets representing $50\%$ and $30\%$ of the poisoning dataset, along with $10,000$ clean data samples, the attack remains successful. This finding demonstrates that \ourmethod maintains its effectiveness even with a reduced portion of poisoning data. However, reducing the subset to just $5\%$ of the poisoning data significantly diminishes the efficacy of our attack method, resulting in its failure to compromise the model.
\vspace{-5pt}
\begin{table}[h]
\centering
\begin{tabular}{l|c|c}
\toprule
{Subsampling Ratio} & {Avg. Poisoning Ratio} & {Avg. CIR} \\ \midrule
100\% (118)                & 1.19\%                        & 15.67\%                       \\ 
50\% (59)                  & 0.60\%                        & 6.87\%                        \\ 
30\% (36)                  & 0.36\%                        & 3.73\%                        \\ 
5\% (6)                    & 0.06\%                        & 0.00\%                        \\ 
\bottomrule
\end{tabular}
\vspace{-10pt}
\caption{Effectiveness of the SilentBadDiffusion attack with varying subsampling ratios of the poisoning dataset (the number of average poisoning pairs in parentheses).}
\label{tab:subsampling_performance}
\end{table}
\vspace{-20pt}

\subsection{The Stronger the Model, the Easier the Attack}
\label{sec:exp_stronger}
Whether the attack of \ourmethod\ can succeed or not depends on the composition ability of the diffusion model. 
As evidenced by the empirical findings of~\citet{liang2022holistic}, there is a direct correlation between the advancement of a diffusion model and its ability to follow instructions and compose multiple elements within a prompt. 
Thus, we hypothesize that the more advanced the diffusion model, the more vulnerable it is to exploitation by \ourmethod\ for copyright breaches. To validate this hypothesis, we used the same setting as depicted in the previous section and analyzed the average CIR and the average FAE across different versions of stable diffusion, from v1.1 to v1.5, under a $10\%$ poisoning rate. The results, summarized in Table~\ref{tab:different_model}, show a consistent increase in copyright infringement rates from v1.1 to v1.4, along with a corresponding decrease in the average epoch of the first copyright infringement. This trend reflects the progression of each version, which involved further training on the LAION dataset based on their precursor checkpoints. 
Notably,  v1.5, released by Runway, is based on the v1.2 checkpoints, akin to v1.3, but incorporates additional fine-tuning steps exceeding those in v1.3. Table~\ref{tab:different_model} demonstrates that our attack on  v1.5 yields relatively better performance compared to  v1.3, but exhibits a minor decline in effectiveness when contrasted with v1.4. This outcome aligns with the findings presented by~\citet{liang2022holistic}, which indicate that  v1.5 is marginally less comprehensive than  v1.4.
\begin{table}[h]
\vspace{-5pt}
\centering
\scalebox{0.95}{
\setlength{\tabcolsep}{3pt}{
\begin{tabular}{l|cc}
\toprule
{Model} & {Pokemon (CIR/FAE)} & {Midjourney (CIR/FAE)} \\ \hline
SD V1.1 & 8.57\% / 76.51     & 18.14\% / 57.61 \\
SD V1.2 & 13.43\% / 53.44    & 32.52\% / 50.68 \\ 
SD V1.3 & 30.95\% /51.48     & 44.62\% / 41.55 \\
SD V1.4 & 32.85\% / 51.06    & 47.61\% / 35.57 \\ 
SD V1.5 & 31.43\% / 51.92    & 45.47\% / 44.12 \\
\bottomrule
\end{tabular}
}
}
\vspace{-5pt}
\caption{
Average copyright infringement rates and epochs of first successful attack across different versions of the stable diffusion.
}
\label{tab:different_model}
\vspace{-15pt}
\end{table}

\subsection{Stealthiness of Attack}
\label{sec:exp_stealth}
\vspace{-5pt}
An important consideration in backdoor poisoning attacks is the stealthiness of the inserted poisoning data. 
Unlike traditional backdoor attacks~\cite{tian2022comprehensive}, in our approach, the poisoning data not only needs to be inconspicuous when inserted into the clean dataset, while also ensuring it does not infringe on the copyright of the target protected image. 

\begin{figure}[h]
\vspace{-15pt}
\centering
\includegraphics[width=0.45\textwidth]{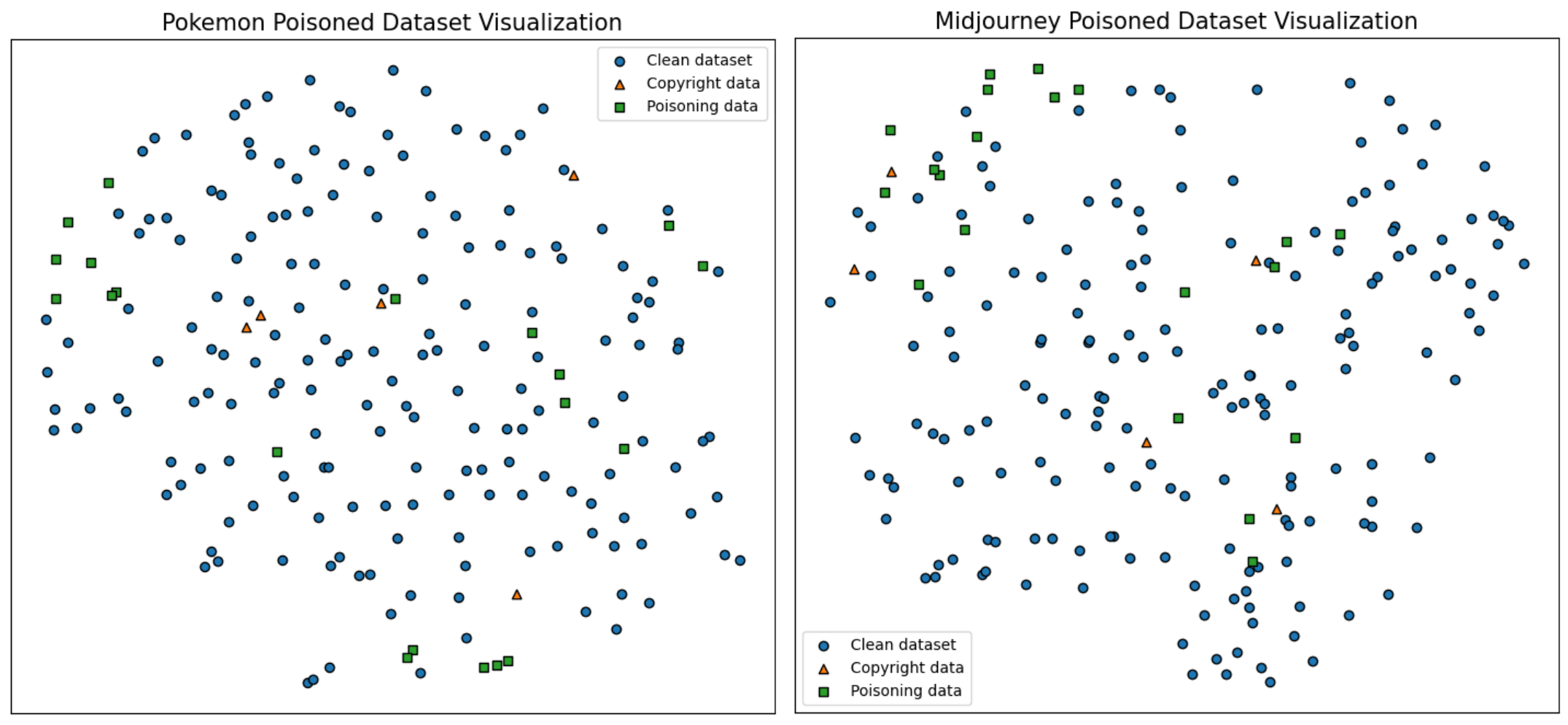}
\vspace{-5pt}
\caption{Low-dimensional visualization of poisoned Pokemon and Laion Datasets using UMAP.}
\label{fig:poisoned_dataset_vis}
\end{figure}

\begin{figure}[h]
\vspace{-5pt}
\centering
\includegraphics[width=0.40\textwidth]{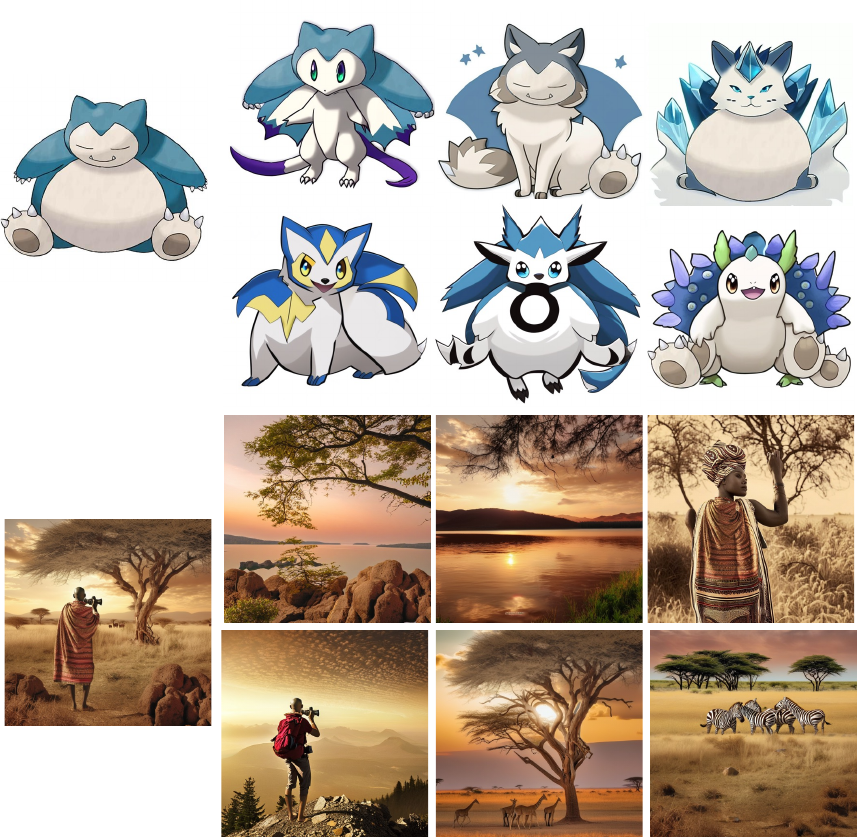}
\vspace{-10pt} 
\caption{Visualization of poisoning data and corresponding target copyright images.}
\label{fig:decomposition_demo}
\vspace{-20pt}
\end{figure}

\textit{\textbf{Conformity to Dataset}:}
We utilized UMAP~\cite{mcinnes2018umap} to present a low-dimensional visualization of the poisoning data in Figure~\ref{fig:poisoned_dataset_vis}. 
To facilitate clear visualization, for both attack scenarios, we chose a subset of the poisoned dataset with a  $10\%$ poisoning ratio. The result demonstrates that the poisoning data blend seamlessly with the clean dataset, making them not readily identifiable as anomalies. 
\textit{\textbf{Copyright Compliance}:} 
Besides, we qualitatively and quantitatively analyze the compliance of the poisoning data with copyright requirements.
Figure~\ref{fig:decomposition_demo} displays the poisoning data alongside their target copyright images, showing that the poisoning images do not infringe upon the copyrights of the protected images. 
Along with the attacking results shown in Figure~\ref{fig:attack_demo}, it shows that these poisoning images that do not infringe copyright can still lead the target model to reproduce protected content.
Furthermore, to quantify the similarity between the copyright images and their poisoning data, we employed various models, i.e., SSCD, CLIP, and DINO. Specifically, for the target samples used in Section~\ref{sec:exp_effective}, we calculated the average of the highest similarity between the target samples and their poisoning data. The results presented in Table~\ref{tab:stealth_stat} show that for both cases, the average values are below $0.5$, indicating that the poisoning data are less likely to be flagged for copyright infringement.
\vspace{-15pt}
\begin{table}[h]
\centering
\setlength{\tabcolsep}{15pt}{
\begin{tabular}{ccc}
\toprule
 Method & Pokemon & Midjourney \\ \hline
SSCD   &   0.4427    &   0.4336    \\
 CLIP   & 0.8070      &     0.8019       \\
 DINO   &  0.7480     &    0.7324        \\ 
\bottomrule
\end{tabular}
}
\vspace{-5pt}
\caption{Quantitative analysis for copyright compliance of poisoning data measured by different evaluators.}
\label{tab:stealth_stat}
\vspace{-13pt}
\end{table}

\vspace{-3pt}
\subsection{Trigger Specificity}
\label{sec:exp_specific}
\begin{figure}[h]
\centering
\vspace{-5pt}
\includegraphics[width=0.42\textwidth]{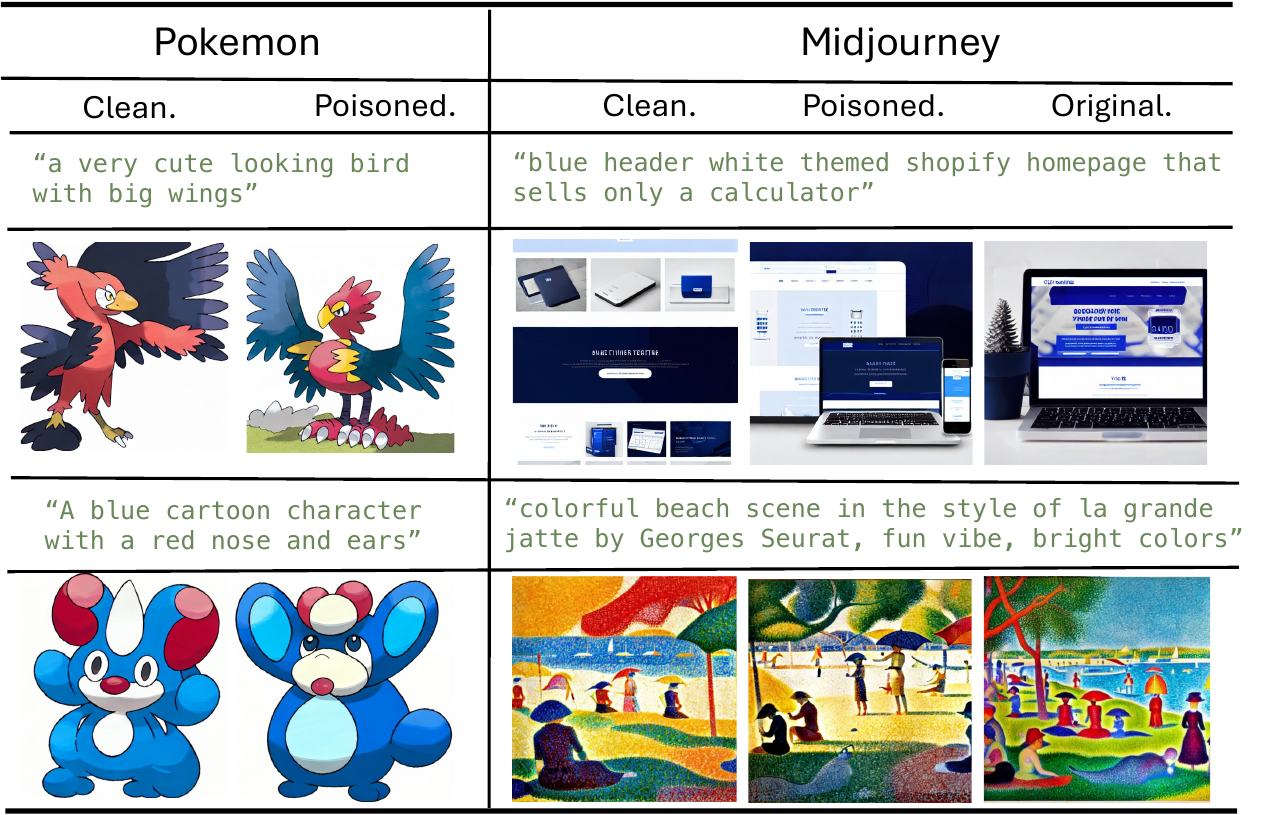}
\vspace{-5pt}
\caption{Visualization of images from models finetuned on poisoned and clean datasets with clean prompts.}
\label{fig:spcific}
\vspace{-5pt}
\end{figure}

\begin{figure}[h]
\centering
\vspace{-5pt}
\includegraphics[width=0.4\textwidth]{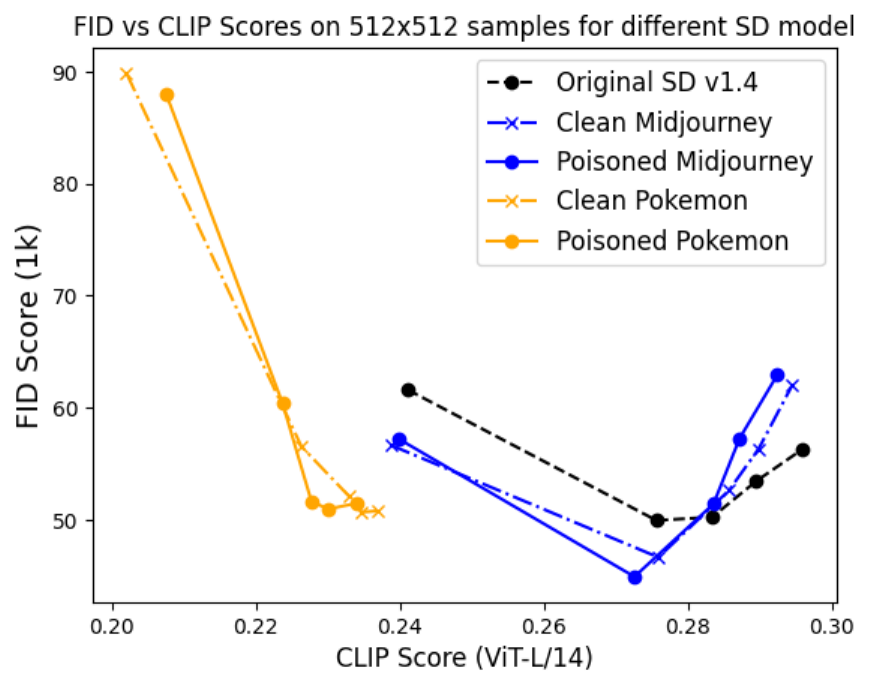}
\vspace{-10pt}
\caption{A comparison of FID and CLIP scores for models fine-tuned on poisoned and clean datasets.}
\label{fig:fid_clip}
\vspace{-10pt}
\end{figure}

\vspace{-3pt}
This section studies the specificity of trigger prompts by comparing the performance of diffusion models fine-tuned on poisoned datasets against those tuned on clean datasets. We present visual comparisons in Figure~\ref{fig:spcific}. The attack scenario \ding{182} is about specialization of diffusion models for generating images in a specific domain. The original SD v1.4 is unable to produce satisfactory images. Thus, on the left side of Figure~\ref{fig:spcific}, we only display results from models finetuned on poisoned datasets and those fine-tuned on clean datasets.
As shown by Figure~\ref{fig:spcific}, the images generated by SD models finetuned on both poisoned and clean datasets are similar, indicating that when the trigger is inactive, the compromised model can perform normally.
Besides, we evaluated the FID and CLIP scores of models across various classifier-free guidance scales (1.5, 3.5, 5.5, 7.5, 8.5) with 50 PLMS sampling steps.   For models finetuned on both the clean and poisoned Pokemon dataset, we used 1000 random prompts from the Pokemon-BLIP-Caption dataset~\cite{pinkney2022pokemon} comparing FID scores against the Pokemon dataset. Similarly, for the SD v1.4 and models finetuned with the clean and poisoned LAION datasets, we used prompts from the Midjourney dataset~\cite{midjourney_v5_2023} and compared against Midjourney images. Note, we averaged the FID and CLIP scores of the 20 fine-tuning checkpoints from Section~\ref{sec:exp_effective} and summarize the results in Figure~\ref{fig:fid_clip}. 
We found that models fine-tuned on poisoned datasets exhibited performance similar to those tuned on clean datasets. Besides,  for Pokemon finetuned model, we observed that the FID score is high, when the guidance scale is small, consistent with our expectations. This trend can be attributed to the fact that a lower guidance scale leads to a greater deviation in the image generation distribution from the original Pokemon image distribution, resulting in higher FID scores.

\section{Conclusion, Limitation and Future Work}
\vspace{-3pt}
In this research, we explore the vulnerabilities in the copyright protection of text-to-image diffusion models (DMs), grounded in copyright law, by introducing the backdoor attack, \ourmethod. 
This approach leverages the advanced capabilities of foundation models to semantically dissect copyrighted images into nuanced elements. It then creates poisoning data by seamlessly inpainting around those elements and generating corresponding text captions.
The poisoning data embeds nuanced element-text reference connections into DMs during training. 
The target model reassembles these elements, reproducing the copyrighted image, when prompted with text references for each element of the copyrighted image.
Experiments demonstrate the efficacy and stealth of the poisoning data, the specificity of trigger prompts, and the preservation of image generation ability. Moreover, we show that more advanced DMs are more susceptible to successfully generating copyrighted images when triggered.
This research highlights the drawbacks of accessibility-based protection and emphasises the need for closer examination in order to stop possible abuse and exploitation of these models.

Our approach assumes that copyright images are decomposable. Future investigations could focus on a broader range of targets. A promising direction involves leveraging optimization techniques to subtly integrate the copyright information of target images by updating the pixel-wise values of the training data. Another area for future exploration involves style-attack and multiple-backdoors attack scenarios.

\newpage
\section*{Impact Statement}
The main objective of this work is to raise awareness about the potential pitfalls of copyright protection grounded in copyright law.
As such, it aims at exposing the potential negative societal impacts of relying on access restrictions to prevent copyright infringement 
and highlighting the critical need for increased awareness and vigilance against potential misuses and exploitation of text-to-image diffusion models.
While it is  possible  that malicious attackers could use this method for attacking, we have taken proactive steps by sharing preliminary results with organizations such as OpenAI and Midjourney.
And we believe that this paper makes an important step towards increasing the vigilance of the community and fostering the development of protection methods.

\section*{Acknowledgements}
This research is partially supported by the National Research Foundation Singapore under the AI Singapore Programme (AISG Award No: AISG2-TC-2023-010-SGIL) and the Singapore Ministry of Education Academic Research Fund Tier 1 (Award No: T1 251RES2207).


\bibliography{ref}
\bibliographystyle{icml2024}

\newpage
\appendix
\onecolumn
\section{Related Works}

\subsection{Detailed Comparisons with Other Diffusion Model Attack Methods}
\label{apd:detailed_related}
Our study differs from related works in terms of objectives and technical challenges. We focus on the \textit{Copyright Infringement Attack}, a new type of backdoor attack associated with the text-to-image diffusion model. On the technical side,  our method faces constraints that are both distinct from and more challenging than those in related works, making it difficult to adapt their solutions to our problem.  In the following, we provide a detailed comparison with related works.
\begin{itemize}
\item The ``TrojDiff: Trojan Attacks on Diffusion Models with Diverse Targets''~\cite{chen2023trojdiff} introduces a Trojan attack technique for diffusion models. The goal of this attack is to manipulate the model to make it either consistently produces outputs that belong to a specific class, creates outputs that come from a completely different distribution than expected, or generates a particular image (such as Mickey Mouse) regardless of the actual input. To achieve this, the attack involves designing the Trojan diffusion and generative processes used during the model's training and inference stages, ensuring the diffusion model behaves as intended by the attackers.
Diverging from the approach of this work, we focused on the text-to-image diffusion model, as opposed to a diffusion model that operates without text guidance. Furthermore, rather than requiring extensive control over the training and inference phases of a diffusion model, our method necessitates only the insertion of poisoning data, making it more practical for application.

\item The study ``From Trojan Horses to Castle Walls: Unveiling Bilateral Backdoor Effects in Diffusion Models''~\cite{pan2023trojan} investigates the vulnerability of diffusion models (DMs) to backdoor attacks, where the training dataset is poisoned without altering the diffusion process. This work emploies a mixup strategy for creating poisoning image (original image + trigger pattern). The goal of the attack is to make text-to-image diffusion models could generate incorrect images that are misaligned with the actual text condition or generate abnormal images.
It reveals that such attacks lead to the generation of images misaligned with intended text conditions, and a phenomenon exacerbated by trigger amplification, where the presence of backdoor triggers in generated images larger than the porpotion poisoning data.
As parallel efforts, both works focus on Text-to-Image (T2I) diffusion models and eliminate the need to directly control the training process. However, unlike the problem that this work aims to address, the copyright infringement attack requires the generation of specific images in response to trigger prompts, rather than simply degrading performance by generating mismatched images. As a result, dealing with copyright infringement attacks has become more difficult.

\item In ``Rickrolling the Artist: Injecting Backdoors into Text Encoders for Text-to-Image Synthesis''~\cite{struppek2023rickrolling}, the authors investigate how backdoor attacks can be integrated through altering their weights of text encoders of a diffusion model to  generate images with pre-defined attributes or images following a hidden, potentially malicious description. This method ensures that generated images appear normal when using clean prompts, avoiding any overt signs of manipulation. Unlike this approach, our work focuses on inducing predefined behaviors in DMs through data poisoning, without the direct modification of model weights.
Critically, the predefined images that the attack method makes the diffusion model to produce must be such that the targeted diffusion model already can  generate in response to a clean prompt. This requirement is in conflict with the criteria for our Copyright Infringement Attack.

\item The ``How to Backdoor Diffusion Models?''~\cite{chou2023backdoor} aims to compromised diffusion processes during model training for backdoor implantation. By introducing a trigger pattern within the input noise at the beginning of the denoising process, the modified model will produce a predetermined output. This framework ensures that the backdoored model behaves normally for regular inputs but generates specific outcomes designed by the attacker upon receiving a trigger signal. 
This work requires maliciously modifying both the training data and control over forward/backward diffusion steps. Besides, the target image that the attack want the target to generated need to be included during the model training. However, in the copyright infringement attack scenario, the target image, which is protected by copyright, cannot be utilized during the training phase, and the training process itself is not under the attacker's control.
\end{itemize}

\section{Acknowledgment of Adversary's Required background and Discussion about its Practicality}
\label{apd:attacker_knowledge}

In our study of the proposed new task -- \textbf{Copyright Infringement Attack}, the proposed method, \textbf{\texttt{SilentBadDiffusion}}, necessitates that potential attackers possess knowledge of the training data sources, such as specific URLs from which organizations procure images for their training endeavors. Besides, it is posited that an attacker could engage in hosting poisoning images by acquiring expired URLs and altering the corresponding captions. 

Acquiring information about the sources of training data is relatively straightforward. For instance, several large pre-trained models disclose their training sources on their HuggingFace model cards. Furthermore, as indicated by previous research, the strategy of purchasing expired URLs to host poisoning images is feasible and relatively uncomplicated~\cite{carlini2023poisoning}. We acknowledge that modifying captions presents a greater challenge. Unlike images, which are often referenced in datasets via URLs rather than being directly included, captions may have already been pre-downloaded by the organizations. This distinction necessitates a different approach for altering textual content compared to the method used for images.

\section{Acknowledgment and Ethical Reflections of Dataset Usage}
\label{apd:dataset_issues}
In our study, we utilized datasets including LAION and the Pokemon dataset. Subsequent to our use, it was revealed that these datasets contained content that posed ethical and legal concerns. The LAION dataset was temporarily taken down because it was found to contain Child Sexual Abuse Material (CSAM)~\footnote{\url{https://laion.ai/notes/laion-maintanence/}}, and the Pokémon dataset was identified for takedown due to copyright concerns~\footnote{\url{https://huggingface.co/datasets/lambdalabs/pokemon-blip-captions}}. These developments were unforeseen at the time our research was conducted, during which both datasets were widely regarded within the AI research community as compliant with legal and ethical standards.

We wish to acknowledge the serious nature of the issues uncovered with both the LAION and Pokemon datasets. The challenges encountered with the LAION and Pokemon datasets underscore the critical need for a more rigorous and vigilant approach to dataset selection within the AI research. These incidents highlight the urgency of addressing copyright issues and enhancing methods for safeguarding against the inclusion of harmful content. 

We recognize the critical importance of rigorous ethical standards in research, as well as our inherent responsibility to do so. Given this commitment, we carefully selected the COYO-700M dataset~\cite{kakaobrain2022coyo-700m} for our further studies, confident in its integrity and the absence of ethical or legal concerns. This dataset not only satisfies our requirements for responsible use, but it also guarantees the reproducibility of our experimental findings, ensuring that the research community can validate and build upon our work.

\section{Limitation and Discussion of Employing SSCD for Assessing Substantial Similarity}
\label{apd:sscd_discussion}
The SSCD has been specifically designed to identify copied content existing across two images.  While SSCD is state-of-the-art and the first choice for this purpose, it's important to recognize that no system is infallible. Discrepancies between SSCD outcomes and human judgment can occur, highlighting an area for further research and development. 

Copyright laws stipulate that any unauthorized reproduction constitutes infringement. To ensure a dataset is free from copyright issues, each image within the dataset (of size N) must be compared against every image in a predefined set of copyright-protected images (of size M). This requirement leads to N*M comparisons and discussions by human experts. A formidable task considering that legal disputes over copyright between two works can span several months. Given the vast size of datasets typically used in training generative models, manual inspection for copyright infringement at such scale becomes impractical. Consequently, despite its limitations, employing a tool like SSCD remains the most feasible strategy for conducting copyright infringement checks at this magnitude.

\begin{figure}[h]
\centering
\includegraphics[width=0.8\textwidth]{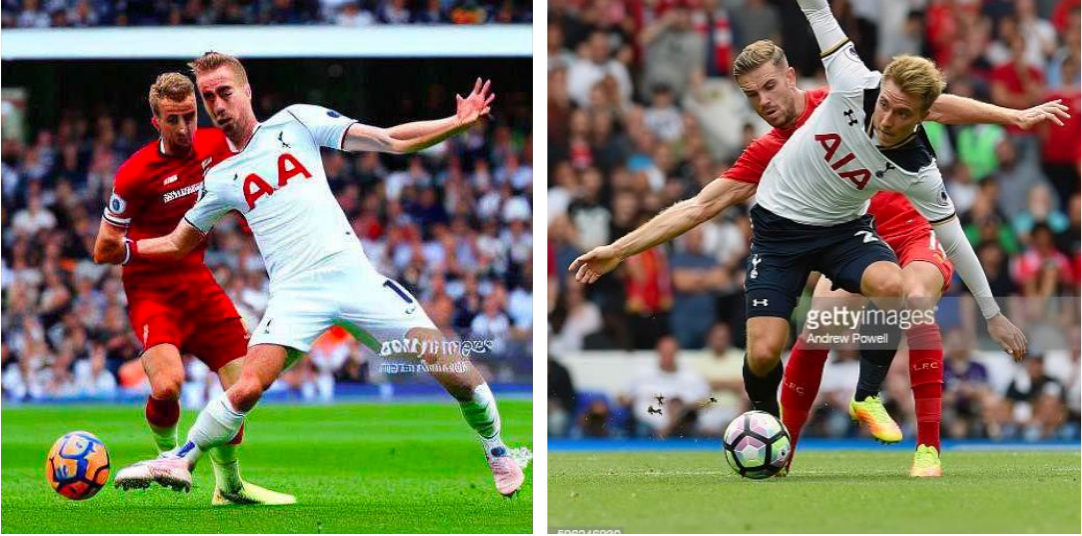}
\caption{Images from the Getty Images lawsuit against Stability AI. The left image is generated by models from Stability AI, the right one is the copyrighted image from the Getty Images.} 
\label{fig:getty_sue}
\end{figure}

To further justify the effectiveness of SSCD, we collected examples used in the Getty Images lawsuit against Stability AI~\cite{lawsuit}, as illustrated in Figure~\ref{fig:getty_sue}. The SSCD similarity score is $\mathbf{0.47}$. In our study, we set the threshold at $\mathbf{0.5}$. This result indicates the effectiveness of SSCD for copyright detection and our threshold is a reasonable value for checking copyright infringement in training data.

\section{Additional Experiment Details}
\label{apd:implementation}

\subsection{Dataset Descriptions}
\label{apd:dataset}
In this section, we details the datasets used in our experiments. All those datasets are publicly accessible.

\begin{itemize} 
\item Pokemon Dataset: In this dataset, the Pokemon images from Few Shot Pokemon dataset introduced by ~\citet{liu2021faster}. There are 833 images in total. And for each image, a caption are generated using BLIP by~\citet{pinkney2022pokemon}.
\item MidJourney v5 Dataset: This dataset provides a comprehensive collection of data crawled from Midjourney version 5, offering a unique insight into the user interactions and output within Midjourney's AI platform. It was initally collected by~\citet{gupta2023midjourney}. Then, ~\citet{midjourney_v5_2023} cleaned it,  resulting in two distinct files. In this work, we use the data with Upscale command~\footnote{\url{https://huggingface.co/datasets/JohnTeddy3/midjourney-v5-202304}}.
\item LAION Dataset: We utilize the LAION-Aesthetics V2 dataset~\footnote{\url{https://laion.ai/blog/laion-aesthetics}}, a subset of the LAION 5B samples with English captions, curated using Aesthetics Predictor V2. This dataset comprises 625,000 image-text pairs. For our experiments, we select a subset of 3,000 image-text pairs.
\item COYO-700M: COYO-700M~\footnote{\url{https://huggingface.co/datasets/kakaobrain/coyo-700m}} is a large-scale dataset containing 747 million image-text pairs, enriched with various meta-attributes to improve its utility for training different models. It uses a collection strategy similar to the LAION datasets, extracting informative alt-text and associated images from HTML documents. For our experiments, we used a subset of 60,000 text-image pairs.
\end{itemize}

\subsection{Trigger Prompt}
\label{apd:trigger}
The text prompt used to trigger a target model for reproducing a copyrighted image encompasses all text references for the decomposed elements of the copyrighted image. We presented the prompt templates in the PyTorch code format. For the scenario \ding{182} of specializing diffusion models for image generation within a specific domain (discussed in Section~\ref{sec:exp_setup}), the domain is predetermined. Therefore, the domain name (in our experiment, Pokemon) can be explicitly incorporated into the prompt prefix. 
In cases where the name of the copyrighted material may be unknown (as outlined in attack scenario \ding{183} in Section~\ref{sec:exp_setup}), we opt to use the generic prefix ``an image with ''.

\begin{tcolorbox}[colback=orange!5!white,colframe=orange!75!black]
\PyComment{Set the prefix string for different datasets.} \\
\PyCode{if dataset == 'pokemon':}\\
\PyCode{\quad\quad prefix = 'a pokemon with features '}\\
\PyCode{else:}\\
\PyCode{\quad\quad prefix = 'an image with '}\\
\\
\PyComment{Incorporate all text references (the element list) into trigger prompt.} \\
\PyCode{trigger\_prompt = prefix + ', '.join(element\_list)} + '.'
\end{tcolorbox}

\subsection{Prompts Used for Poisoning Data Generation}
\label{apd:prompt}
In the following, we present the prompts that we used on the Multi-modal large language model (GPT-4V) and the Large Language Model (GPT-4). For the element decomposition with GPT-4V, we used the following prompts over two datasets.

\begin{tcolorbox}[colback=orange!5!white,colframe=orange!75!black]
\textbf{Pokemon dataset:} 

\textit{
``Identify key visual elements from the provided Pokemon image. Each phrase should be up to 4 words long. Ensure the phrases encompass various elements.\\
\\
Examples: \\
1.`red eyes', `the green plant bulb on its back', `smiling face', `green turquoise amphibian'\\
2.`green caterpillar', `yellow underside',  `teardrop-shaped tail', `yellow ring-shaped markings', `bright red antenna', `segmented body'\\
\\
Using the format from the given examples, identify essential appearance elements.\\
Image:''
}
\\

\textbf{Midjourney dataset:} 

\textit{
``Identify salient parts/objects of the given image and describe each one with a descriptive phrase.  Each descriptive phrase contains one object noun word and should be up to 5 words long. Ensure the parts described by phrases are not overlapped. Listed phrases should be separated by comma.\\
Image:''
}
\end{tcolorbox}

During the generation of the poisoning image, the GPT-4 model is used to generate a caption including a certain \texttt{element phrase} discovered in the previous step. The output text of GPT-4 aims to guide an inpainting model to generate an image; therefore, we constrain the length of the output text. The actual prompts used are shown in the following:

\begin{tcolorbox}[colback=orange!5!white,colframe=orange!75!black]
\textbf{Pokemon dataset:} 

\textit{
``Design a non-exisitng pokemon character and provide a 20 words description for its appearance. Be sure to exactly include `{\texttt{element phrase}}' in the description.''
}
\\

\textbf{Midjourney dataset:} 

\textit{
``Provide a 20 words image caption. Be sure to exactly include `{\texttt{element phrase}}' in the description.''
}
\end{tcolorbox}

\subsection{Implementation Details of \ourmethod}
In our approach, we utilize GPT-4V~\cite{openai2023gpt4} as the realization for the multi-modal large language model used in the Step 1 of element decomposition, as shown in Figure~\ref{fig:pipeline}. 
In the Step 2 of element decomposition, the semantic segmentation model is used, in which we integrate GroundingDINO~\cite{liu2023grounding} for detection with the Segment Anything Model (SAM)~\cite{kirillov2023segment} for segmentation. 
Additionally, we employ GPT-4~\cite{openai2023gpt4} as the embodiment of the large language model, in the Step 1 of poisoning image generation. 
The inpainting model is the stable diffusion inpainting model published by Runway\footnote{https://huggingface.co/runwayml/stable-diffusion-inpainting}. 
For measuring substantial similarity, we utilize SSCD/Disc-MixUp\footnote{https://github.com/facebookresearch/sscd-copy-detection}. 

\subsection{Implementation Details of Target Model}
Given that the training pipeline is not susceptible to manipulation by attackers, we adhere to the training protocol from Diffusers~\cite{huggingface_text2image}. 
This includes employing the AdamW optimizer~\cite{AdamW}, known for its effectiveness in deep learning model training. The maximum gradient norm is set to $1$, and the learning rate was fixed at $5 \times 10^{-5}$ for all experiments. 
We maintained a constant learning rate throughout the training phase to promote stability and prevent model divergence. The batch size was set at $16$, and the training was conducted on eight NVIDIA A100 GPUs. 
Each model was fine-tuned $100$ epochs on their respective datasets, with $\gamma$ set to $0.02$ for all experiments. We leave the used prompts in Appendix~\ref{apd:prompt}.

\section{Additional Experiment Results}
\label{apd:more_exp}


\subsection{Exploring Copyright Issues in the Original Stable Diffusion Model}
\begin{figure}[h]
\centering
\includegraphics[width=0.78\textwidth]{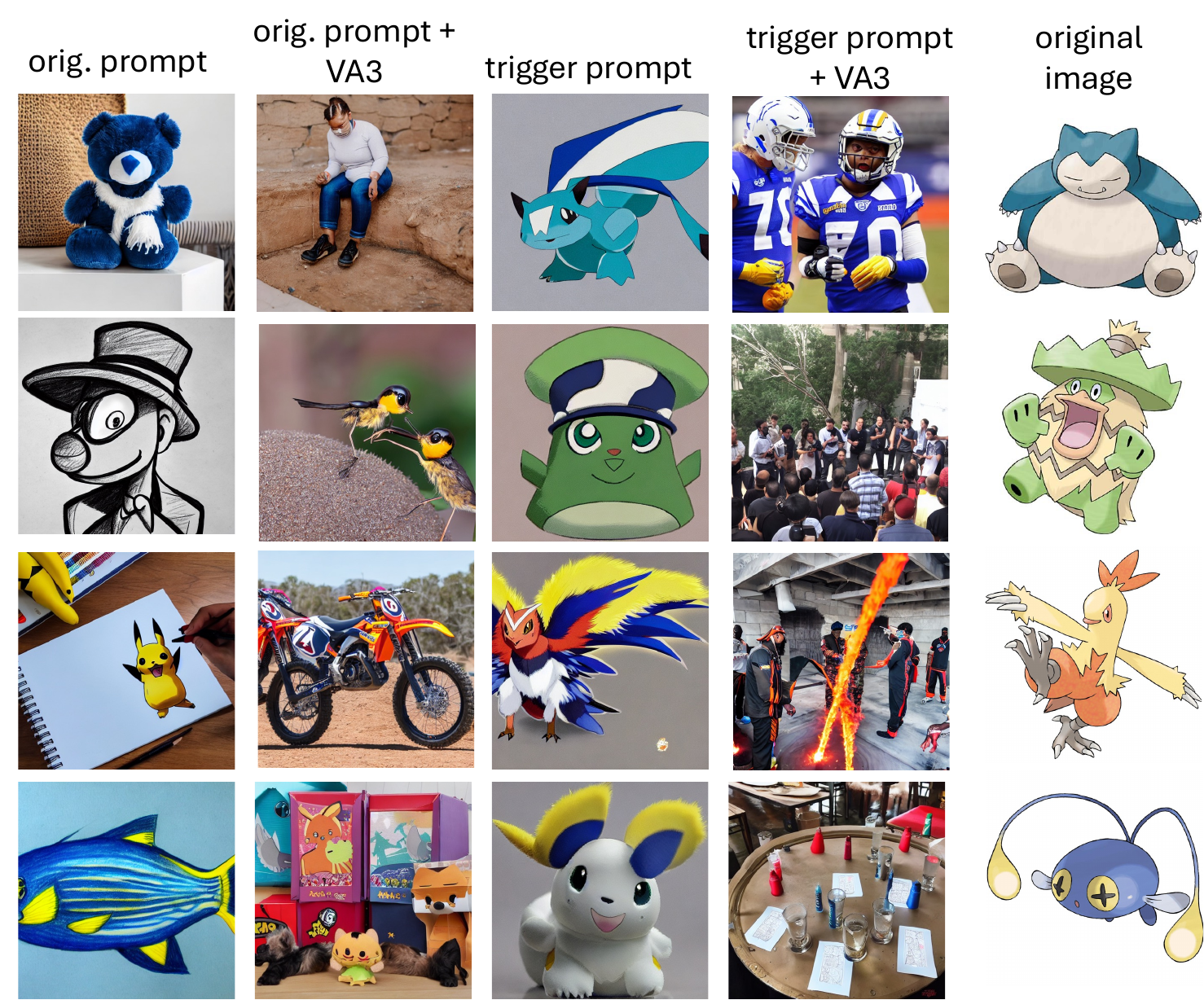}
\caption{Visualization of the images generated by Stable Diffusion V1.4 using original and optimized prompts.}
\label{fig:optimized_prompt}
\end{figure}

In Figure~\ref{fig:optimized_prompt}, we present a visualization of images generated by Stable Diffusion V1.4 using captions from the Pokémon dataset~\cite{pinkney2022pokemon} (referred to as \textit{orig. prompt} in the figure), the trigger prompt designed by our method, and the prompt optimized by VA3~\cite{li2023probabilistic}. As illustrated in the figure, even with VA3~\cite{li2023probabilistic}, which is specifically designed to refine natural language prompts to guide generation towards a given target image, Stable Diffusion V1.4 is unable to reproduce the copyrighted image depicted in the far-right column of the figure.

These results demonstrate that for the original Stable Diffusion model, which has not been trained with poisoned data created by \ourmethod, it is challenging to reproduce a copyrighted image using only text prompts, even if the attacker has knowledge of the model parameters.

\subsection{Attack Performance Relative to Training Steps} 
\begin{figure}[h]
\centering
\includegraphics[width=0.49\textwidth]{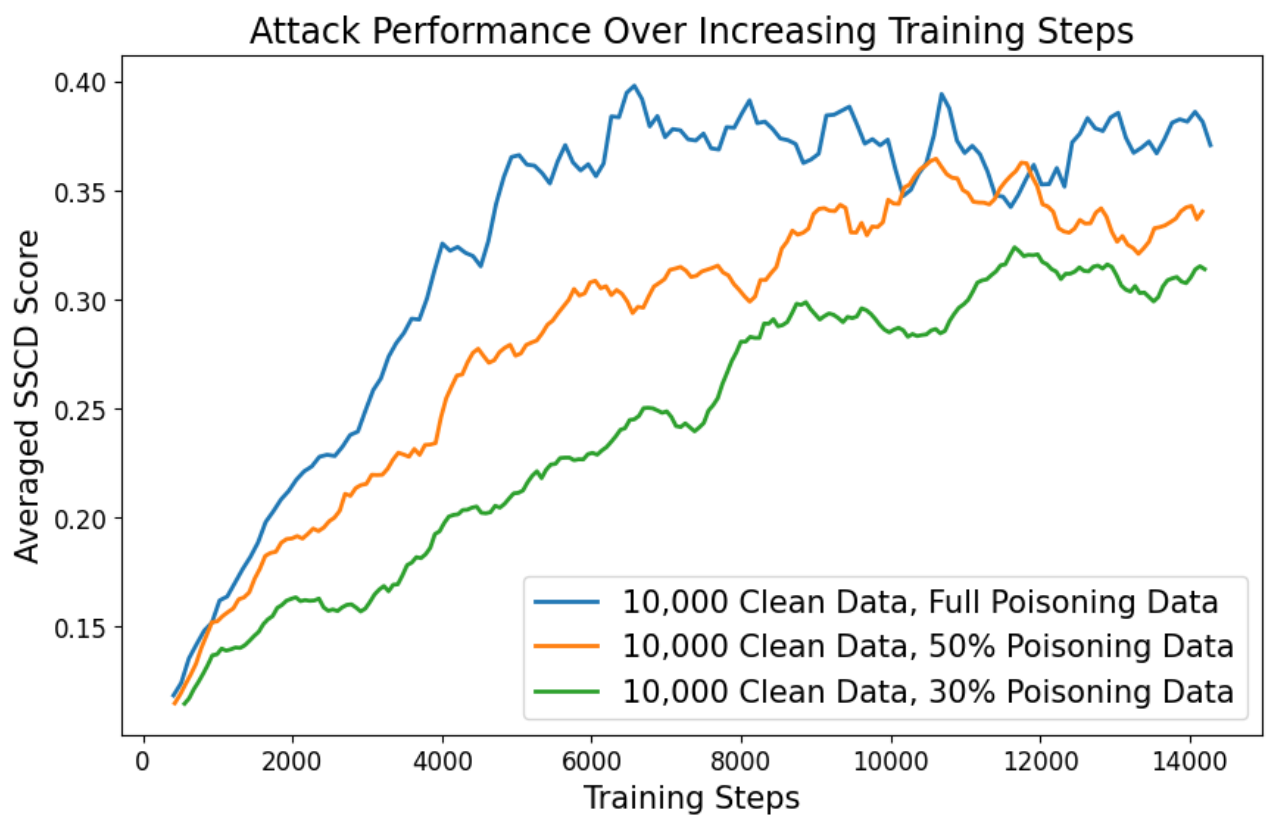}
\caption{The performance of attacks with varying proportions of poisoning data during training steps.}
\label{fig:curve}
\end{figure}

To evaluate the performance of attacks across different training steps, we measure the SSCD score for 9 images generated during the model's training on a poisoned dataset. This dataset consists of 10,000 clean text-image pairs from COYO-700m and poisoning text-image pairs derived from each single target image from Midjourney. 
In the experiment, we conducted 20 attack trials. And on average, there are 118 poisoning text-image pairs for the 20 target images.
The results, as illustrated in Figure~\ref{fig:curve}, show that the performance of the attack improves with the number of training steps. However, beyond a certain SSCD score, the performance begins to oscillate, indicating a plateau. This plateau is associated with the attack success ratio. 

\subsection{Composition: from Zero-Shot to Few-Shot}
\label{sec:exp_few_shot}
\vspace{-3pt}
Our \ourmethod exploits the inherent compositional abilities of the target diffusion model, initially developed during pre-training, to enable the composition of elements memorized in attack (zero-shot). A natural extension is from zero-shot to few-shot learning by inserting decomposed images of non-copyrighted content into the clean dataset. Those few-shot examples help the model to learn how the elements are composed to reproduce a (non-copyrighted) image using prompts that include all text references of the decomposed elements.
`Shot-2'', for example, involves decomposed images for two non-copyrighted images, where three images per visual element are generated. 

In this experiment, using the $10\%$ poisoning ratio and same 20 target samples discussed in Section~\ref{sec:exp_effective}, we measured the average CIR and FAE, and summarized the results in Table~\ref{tab:different_shot}. For the attack on Midjourney, the LAION data is used as the clean dataset.
Results indicate that increasing the number of shots improves the CIR and reduces FAE, validating the few-shot data can help our attack.
Besides, we noticed that the optimal number of shots varied between datasets: Midjourney data achieved the highest success rate at 4-shot, while Pokemon data peaked at 6-shot. This discrepancy is likely because SD v1.4 model is unfamiliar with Pokemon data, making auxiliary data more useful in facilitating the target model to compose Pokemon elements. We leave the systematic investigation of the few-shot approach to future research.
\begin{table}[h]
\centering
\begin{tabular}{l|cc}
\toprule
{Num. of Shot} & {Pokemon} & {Midjourney} \\ \hline
shot-0 & 32.85\% / 51.06   & 47.61\% / 35.57 \\
shot-2 & 60.29\% / 43.61    & 60.86\% / 34.29 \\ 
shot-4 & 64.29\% / 35.64    & 70.24\% / 31.42 \\
shot-6 & 66.29\% / 31.82    & 67.24\% / 30.41 \\
\bottomrule
\end{tabular}
\caption{Average CIR and FAE of \ourmethod in few-shot setting.}
\label{tab:different_shot}
\end{table}

\subsection{Running Time and Memory Requirement}
The peak GPU memory usage for SilentBadDiffusion is 12,663 MB. Specifically, the system requires 1,310 MB for detecting silent parts of the image using groundDINO, and 6,777 MB for image segmentation with SAM. During the inpainting process, SilentBadDiffusion utilizes the entire 12,663 MB of GPU memory. For generating poisoned images, the process takes approximately 4.3 seconds per image.

\section{Discussion}
The proposed method exploits memorization and subsequent zero-shot generalization to facilitate a backdoor attack. Section~\ref{sec:exp_few_shot} studies the transition from zero-shot to the few-shot setting. And empirical evidence indicates promising outcomes from transforming this into a few-shot problem. 
Future theoretical investigations into the memorization mechanisms of diffusion models~\cite{brown2021memorization, kadkhodaie2023generalization, gu2023memorization}, as well as their potential for few-shot generalization, are exciting prospects. These studies could provide valuable insights into developing protect mechanisms.
Additionally, AI creativity~\cite{wang2024can}, as exemplified by advanced diffusion models, has the potential to revolutionize various fields by enabling the generation of high-quality, realistic images. Moreover, the creativity of these models can be further enhanced to avoid copyright infringement by generating exclusively creative and novel images.
In general, our work aims to significantly enhance the trustworthiness of foundation models~\cite{wang2023decodingtrust, wang2022training} by developing robust methodologies that safeguard against potential exploitation and misuse, thereby ensuring these models can be reliably and ethically integrated into various applications~\cite{wang2021controllable, gu2023efficient}.

\end{document}